\documentclass[prl,amsmath,amssymb,aps,twocolumn]{revtex4-1}

\usepackage{graphicx}
\usepackage{dcolumn}
\usepackage{bm}
\usepackage{epstopdf}
\usepackage{color}

\begin{document}

\preprint{APS/123-QED}

\title{Method of spectral Green functions in driven open quantum dynamics}

\author{A. Karabanov, W. K\"ockenberger}
\affiliation{Sir Peter Mansfield Imaging Centre, School of Physics and Astronomy, University of Nottingham, University Park, Nottingham, NG7 2RD, UK}

\begin{abstract}
A novel method based on spectral Green functions is presented for the simulation of driven open quantum dynamics that can be described by the Lindblad master equation in Liouville density operator space. The method extends the Hilbert space formalism and provides simple algebraic connections between the driven and non-driven dynamics in the spectral frequency domain. The formalism shows remarkable analogies to the use of Green functions in quantum field theory such as the elementary excitation energies and the Dyson self-energy equation. To demonstrate its potential, we apply the  novel method to a coherently driven dissipative ensemble of 2-level systems comprising a single ``active'' subsystem interacting with $N$ ``passive'' subsystems --- a generic model with important applications in quantum optics and dynamic nuclear polarization. The novel method dramatically reduces computational cost compared with simulations based on solving the full master equation, thus making it possible to study and optimize many-body correlated states in the physically realistic limit of an arbitrarily large $N$.

\end{abstract}

\maketitle

{\bf\em Introduction.} 
Open quantum dynamics takes into account the environment (outer degrees of freedom) and so more accurately describes real physical phenomena compared with closed quantum dynamics based entirely on the energy operator (inner Hamiltonian). The environment tends to quench coherent states and to purge quantum information by free thermal decay. On the other hand, it can cause irreversible dynamics that makes it possible to create and keep coherent states by continuous driving the system out of its thermal equilibrium \cite{bp-02,k-08,vwc-09}. This makes driven open systems a fundamental object of quantum theory.  

The mathematical description of the dynamics of open quantum systems is non-unitary and generally complex. This especially refers to correlated many-body systems that exhibit collective phenomena depending on their environment. In many cases the Lindblad master equation approach can be used that retains the positivity of the density operator and introduces the environmental effects through Markovian quantum jump operators that enter the dissipator in a simple algebraic way \cite{l-76,gks-76,bp-02,k-08,vwc-09,w-15}. However, the Liouville space containing the trajectories of the density operator describing the evolution of a many-body open quantum system grows exponentially with the number of constituents and the dynamics is sensitive to a large number of physical parameters. Hence efficient mathematical tools are necessary to perform adequate approximations and state space restrictions in order to gain insight into the underlying physics.

For the description of collective phenomena in large-scale closed quantum systems at thermal equilibrium, a method involving the use of Green functions was developed that statistical physics adopted from quantum field theory \cite{z-60,agd-65,r-14}. This method is based on the approximate calculation of correlations between dynamic operators that leads to self-consistent equations for the observables. Subsequently, this method was extended to non-equilibrium closed and non-Markovian open quantum systems describing various transport phenomena where, besides the standard time domain formalism, it was transformed to an inhomogeneous spectral problem in Hilbert space \cite{d-95,v-14,kk-19,te-18,cg-20}. 

Here we propose an extension of the non-equilibrium spectral approach to the important class of driven Markovian open quantum dynamics in the Liouville space of the density operator. To this end we show that the Green function for an inhomogeneous spectral problem  can be formulated in terms of both Hamiltonian and dissipative parts of the Lindblad master equation. The steady state of the driven system is then obtained by a simple algebraic transform of the non-driven thermal equilibrium. Remarkably there is a close analogy between our proposed spectral formalism and the use of Green functions in quantum field theory, including the elementary excitation energies and the Dyson self-energy equation. For a demonstration, we apply the method to a coherently driven dissipative ensemble of correlated 2-level systems, a basic model used in quantum optics and dynamic nuclear polarization. We show that, because the computational cost of the method is significantly cheaper in comparison with the direct master equation simulation, it is possible to study and optimize the many-body correlated states in the realistically large-scale limit. This opens up new possibilities in simulations of many-body driven open quantum dynamics including the spectral response to the driving and the fast search for the optimal parameter regions.  

{\bf\em Spectral Green functions for the Lindblad master equation.}
The dynamics of open quantum systems is described in terms of the Lindblad master equation \cite{l-76,gks-76} 
\begin{equation}
\dot\rho={\cal M}\rho,\quad
{\cal M}=-i\left[H,\cdot\right]+{\cal D}
\label{me}
\end{equation} 
where $\rho$ is the density operator, $H$ is the Hamiltonian describing the (internal) energy of the system and $\cal D$ is the dissipator that represents the effect of the (external) environment. The latter uses Markovian jumps represented by (dimensionless) operators $X_j$ and system-environment exchange rates $\gamma_j$,
\begin{equation}
{\cal D}=\sum\gamma_j{\cal L}(X_j),\
{\cal L}(X)\rho=X\rho X^\dagger-\frac{1}{2}\{X^\dagger X,\rho\}.
\label{D}
\end{equation} 
The dissipator tends to return the system to the state that is in thermal equilibrium with the environment while the Hamiltonian contains terms that drive the system out of thermal equilibrium, ${\cal D}\rho_{\rm th}=0$, $[H,\rho_{\rm th}]\not=0$. Eq.~(\ref{me}) describes a driven open quantum dynamics in the Liouville space of the density operator. 

The master equation preserves the unit trace ${\rm Tr}\rho=1$. The superoperator $\cal M$ transfers all operators to traceless operators and hence reduces the dimension of the Liouville space. As a consequence, $\cal M$ is degenerate with a non-trivial zero eigenspace. This eigenspace contains a non-thermal steady state that is eventually established in the driven system, 
\begin{equation}
t\to+\infty,\quad
e^{{\cal M}t}\rho_{\rm th}\to\rho,\quad
{\cal M}\rho=0,\quad
{\rm Tr}\rho=1.
\label{mess}
\end{equation}   
In the fully dissipative case the zero eigenspace is 1-dimensional, Eq.~(\ref{mess}) uniquely defines the steady-state for all initial conditions. In this case the only traceless solution to Eq.~(\ref{mess}) is trivial,
\begin{equation}
{\cal M}\rho=0,\quad
{\rm Tr}\rho=0\quad
\longrightarrow\quad
\rho=0.
\label{tr}
\end{equation}

Suppose the Hamiltonian can be represented in the form
\begin{equation}
H=P+H_0+\zeta H_1,\
[H_{0,1},\rho_{\rm th}]=0,\
[P,\rho_{\rm th}]\not=0
\label{H01}
\end{equation}
where $\zeta$ is a real scalar parameter, the Hermitian operators $H_{0,1}$ are non-driving and the Hermitian operator $P$ contains the driving terms of the Hamiltonian. We will assume that the operator $H_1$ is dimensionless and $H_0,\,P$ and $\zeta$ are measured in frequency units. Extracting the thermal equilibrium part $\rho=\rho_{\rm th}+\bar\rho$, ${\rm Tr}\bar\rho=0$ and introducing the superoperators ${\cal F}_0=-i[H_0,\cdot]+{\cal D}$, ${\cal P}=i[P,\cdot]$, ${\cal H}_1=i[H_1,\cdot]$, the homogeneous Eq.~(\ref{mess}) is rewritten as an inhomogeneous generalized spectral problem
\begin{equation}
\left({\cal F}_0-{\cal P}-\zeta{\cal H}_1\right)\bar\rho={\cal P}\rho_{\rm th}
\label{sp}
\end{equation}
where $\zeta$ plays the role of a spectral parameter and the solution $\bar\rho$ belongs to the subspace of traceless operators. 

Assuming the validity of Eq.~(\ref{tr}) for any real value of $\zeta$, the superoperator ${\cal F}_0-{\cal P}-\zeta{\cal H}_1$ is non-degenerate and hence invertible. Then the unique solution to Eq.~(\ref{sp}) is $\bar\rho={\cal G}(\zeta){\cal P}\rho_{\rm th}$ where the superoperator ${\cal G}(\zeta)$ must satisfy the equation
\begin{equation}
\left({\cal F}_0-{\cal P}-\zeta{\cal H}_1\right){\cal G}(\zeta)=1
\label{G}
\end{equation}
with the unit superoperator in the right-hand side. The superoperator ${\cal G}(\zeta)$ is independent of the thermal equilibrium, acts in the subspace of traceless operators and plays the role of the Green function of the inhomogeneous spectral problem (\ref{sp}). We call the superoperator ${\cal G}(\zeta)$ {\it the driven spectral Green function} for Eqs.~(\ref{me}), (\ref{H01}). By virtue of the previous equations, the steady state is written as
\begin{equation}
\rho=\left[1+{\cal X}(\zeta)\right]\rho_{\rm th},\quad
{\cal X}(\zeta)={\cal G}(\zeta){\cal P}.
\label{rhoG}
\end{equation}  
Eq.~(\ref{G}) can be rewritten in the form
\begin{equation}
\left[1-{\cal G}_0{\cal P}\right]{\cal G}(\zeta)={\cal G}_0(\zeta)
\label{GG}
\end{equation}
where the superoperator ${\cal G}_0(\zeta)$ must satisfy the equation
\begin{equation}
\left({\cal F}_0-\zeta{\cal H}_1\right){\cal G}_0(\zeta)=1.
\label{G0}
\end{equation}
Indeed, multiplying both sides of Eq.~(\ref{GG}) by the invertible superoperator ${\cal F}_0-\zeta{\cal H}_1$, we obtain to Eq.~(\ref{G}). Eq.~(\ref{G0}) uniquely defines the superoperator ${\cal G}_0(\zeta)$ that is independent of the driving part $P$ of the Hamiltonian. We will call the superoperator ${\cal G}_0(\zeta)$ {\it the non-driven spectral Green function}. Using Eq.~(\ref{GG}), the solution to Eq.~(\ref{sp}) becomes $\bar\rho={\cal X}(\zeta)\rho_{\rm th}=\left[1-{\cal G}_0(\zeta){\cal P}\right]^{-1}{\cal G}_0(\zeta)
{\cal P}\rho_{\rm th}$. Applying the universal operator relation $1+(1-Y)^{-1}Y=(1-Y)^{-1}$, we obtain then for the steady state
\begin{equation}
\left[1-{\cal X}_0(\zeta)\right]\rho=\rho_{\rm th},\quad
{\cal X}_0(\zeta)={\cal G}_0(\zeta){\cal P}.
\label{rhoG0}
\end{equation}  
The dual Eqs.~(\ref{rhoG}), (\ref{rhoG0}) provide compact formulas for the steady state $\rho$ of the master Eq.~(\ref{me}) as a linear transformation of the thermal equilibrium $\rho_{\rm th}$ defined by the product of the driving superoperator $\cal P$ and the driven and non-driven spectral Green functions ${\cal G},\,{\cal G}_0$ determined by Eqs.~(\ref{G}), (\ref{G0}) and connected via Eq.~(\ref{GG}).  

The superoperator ${\cal F}_0-{\cal P}-\zeta{\cal H}_1$ of Eq.~(\ref{G}) linearly depends on $\zeta$, hence the driven Green function 
${\cal G}(\zeta)$ is rationally extendable into the complex plane of $\zeta$
\begin{equation}
{\cal G}(\zeta)={\cal G}^{(0)}+\sum_{r=1}^m\left(\zeta-\zeta_r\right)^{-1}{\cal G}^{(r)}.
\label{spex}
\end{equation}
Here the poles $\zeta=\zeta_r$ and the residues ${\cal G}^{(r)}$ are given by (suitably normalized) solutions to the homogeneous driven spectral problem $\left({\cal F}_0-{\cal P}-\zeta_r{\cal H}_1\right){\cal G}^{(r)}=0$. The superoperator 
${\cal F}_0-{\cal P}-\zeta{\cal H}_1$ is real and non-degenerate for real $\zeta$, so the poles have nonzero imaginary parts and exist in complex conjugate pairs. By Eq.~(\ref{rhoG}), the superoperator ${\cal X}(\zeta)$ and the steady state have rational expansions with the same poles. Similarly, the non-driven Green function has poles and residues defined by the homogeneous non-driven problem 
$\left({\cal F}_0-\zeta{\cal H}_1\right){\cal G}=0$.

Eqs.~(\ref{G}), (\ref{G0}) can be considered to be a Liouville space extension of the spectral Green function formalism in Hilbert space \cite{d-95, v-14,kk-19}. There are noteworthy analogies to quantum field theory. The real parts of the poles $\zeta=\zeta_r$ of the superoperator ${\cal G}(\zeta)$ in Eq.~(\ref{spex}) play the roles of the elementary excitation energies. Eq.~(\ref{GG}) is a copy of the Dyson equation with the superoperators ${\cal G}_0$, $\cal G$ and $\cal P$ playing the roles of the bare and dressed propagators and the self-energy \cite{z-60,agd-65,r-14}. 

The advantage of the method introduced in this section is that the use of Eqs.~(\ref{G}), (\ref{rhoG}) is generally much less computationally costly than calculating the steady state as the dynamic limit or an element of the zero eigenspace by Eq.~(\ref{mess}). Indeed, the former requires only an operator inversion while the latter needs either calculation of an operator exponent or an operator diagonalization. Besides, the knowledge of the poles and residues of the rational structure (\ref{spex}) and Eq.~(\ref{rhoG}) can be used to evaluate the steady state once for all values of the spectral parameter, thus justifying the importance of the method for spectroscopy. At the poles $\zeta=\zeta_r$ the superoperator ${\cal X}(\zeta)$ becomes infinite. Hence, the real values of the spectral parameter closest to the poles $\zeta\sim{\rm Re}\,\zeta_r$ define the spectral peaks, i.e., the physical regions where the maximal response of the system to the driving should be expected. The imaginary parts ${\rm Im}\,\zeta_r$ give the Lorentzian widths of the spectral peaks. Furthermore,  Eqs.~(\ref{rhoG}), (\ref{rhoG0}) imply that the superoperators $1+{\cal X}(\zeta)$, $1-{\cal X}_0(\zeta)$ are inverse to each other and so ${\cal X}(\zeta)$, ${\cal X}_0(\zeta)$ commute for all $\zeta$ and are diagonalized in the same basis. Eq.~(\ref{rhoG0}) admits the formal expansion (convergent for ${\cal X}_0$ close to nilpotent)
\begin{equation}
\rho=\left[1+{\cal X}_0(\zeta)+{\cal X}_0^2(\zeta)+\ldots\right]\rho_{\rm th}.
\label{exp}
\end{equation} 
providing the zero, linear, quadratic, etc, responses of the steady state to the driving. It follows from Eq.~(\ref{exp}) that the poles of the non-driven Green function are also poles of the driven Green function. Eq.~(\ref{rhoG0}) implies that the latter has extra poles defined by the scalar equation ${\rm det}\,\left[1-{\cal X}_0(\zeta)\right]=0$ extracting those values of $\zeta$ where the superoperator ${\cal X}_0(\zeta)$ has a non-trivial fixed point (an eigenoperator with the unit eigenvalue) 
${\cal X}_0(\zeta)\rho=\rho$. 

The polynomial resolution (``renormalization'') of the perturbation series of Eq.~(\ref{exp}) as well as the important links of the spectral Green functions to the time-domain Green functions \cite{z-60,agd-65,r-14} and the projection methods \cite{kSE-12,kCE-12,k-15,k-18} are given in Appendix, A. 

{\bf\em Application to ensemble of 2-level systems.}

We now illustrate the method of spectral Green functions by its application to a driven dissipative ensemble of correlated 2-level quantum systems --- the generic model system to study collective phenomena in quantum optics, magnetism and quantum information \cite{l-73,nr-09,nc-10}.     

The model Hamiltonian of Eq.~(\ref{me}) that we will consider is built of $N+1$ correlated 2-level quantum systems comprising 
one ``active'' subsystem described by the spin-1/2 angular momentum ${\bf S}$ and $N$ ``passive'' subsystems characterized by spin-1/2 angular momenta ${\bf I}^{(k)}$, $1\le k\le N$, featuring the non-driven and driving parts of Eq.~(\ref{H01})
\begin{equation}
H_0=0,\quad
H_1=S_z,\quad
P=\Omega\left(I_+S_-+I_-S_+\right).
\label{Hm}
\end{equation}
Here ${\bf I}=\sum_k{\bf I}^{(k)}$ is the total passive spin and $\Omega$ is the effective driving strength. To describe the model dissipator of Eqs.~(\ref{me}), (\ref{D}), we approximate the thermal equilibrium density operator as $\rho_{\rm th}=(N+1)^{-1}(1-2S_z)$, so that the active subsystem is in the ground state while the passive subsystems have all equally populated levels. The dissipation is built of separate active and passive parts ${\cal D}={\cal D}_S+{\cal D}_I$ written in the Lindblad form
\begin{equation}
\begin{array}{c}
{\cal D}_S=\Gamma_1{\cal L}(S_-)+2\Gamma_2{\cal L}(S_z),\\[2mm]
{\displaystyle{\cal D}_I=\frac{\gamma_1}{2}\left[{\cal L}(I_+)+{\cal L}(I_-)\right]+2\gamma_2{\cal L}(I_z).}
\end{array}
\label{Dm}
\end{equation}       
Here $\Gamma_1,\,\Gamma_2,\,\gamma_1,\,\gamma_2>0$ are the effective active and passive longitudinal and transverse relaxation rates. 

The chosen model has two important applications. In quantum optics, the active subsystem describes the pumped (solid atomic or molecular) gain medium while the passive ensemble plays the role of the population inverted amplifier \cite{s-98,a-74}. In high field solid state dynamic nuclear polarization, the active subsystem is formed by a microwave irradiated unpaired electron spin (of a free radical or paramagnetic ion) while the passive subsystems belong to nuclear spins in the proximity of the electron \cite{a-61,wb-07}. The driving is caused by a time-periodic (optical or microwave) excitation and the rotating wave approximation and an effective Hamiltonian must be applied by a suitable transformation \cite{kSE-12,kCE-12} and averaging over the passive ensemble. It is assumed that the passive dissipation is dominated by the collective relaxation mechanisms \cite{a-74,bzpp-69,b-49,ag-78}. The details of the derivation and physical conditions of the model are given in Appendix, B. 

The Hamiltonian and dissipative parts are invariant to permutations of the passive subsystems and depend only on the components of the total passive spin, so the passive subsystems are identical and it is sufficient to represent them by a single angular momentum $\bf I$ with the spin quantum number $I=N/2$, similar to the Dicke model \cite{d-54,bzpp-69,g-11,lph-81,dc-78}. The corresponding occupation numbers are defined as $n=(n_+-n_-)/2=-I,\,-I+1,\,\ldots,\,I$ where $n_\pm$ are the numbers of passive subsystems in the excited/ground state. The master equation with the Hamiltonian and dissipative parts defined by Eqs.~(\ref{Hm}), (\ref{Dm}) preserves the subspace $\Lambda_0$ of zero-quantum coherences, $[I_z+S_z,\Lambda_0]=0$. Since $\rho_{\rm th}\in\Lambda_0$, the driven dynamics and the steady state are closed in $\Lambda_0$. The density operator has the representation
\begin{equation}
\begin{array}{c}
\rho=\rho^{(0)}+\rho^{(1)},\quad
\rho^{(1)}=\rho_+S_-+\rho_-S_+,\\[2mm]
\rho^{(0)}=\rho_0\left(1/2-S_z\right)+2\rho_zS_z
\end{array}
\label{0}
\end{equation} 
with $\rho_{0,z,\pm}$ containing only the zero-quantum $[I_z,\rho_{0,z}]=0$ and single-quantum $[I_z,\rho_\pm]=\pm\rho_\pm$ coherences of the passive spin ensemble. The operators $\rho_{0,z}$ are diagonal in the basis generated by the occupation numbers, while the operators $\rho_\pm$ are represented by the upper and lower secondary diagonal matrices. The total dimension of the problem equals then $2(2I+1)+2\cdot 2I=4N+2$. Direct numerical solving in terms of the full master equation involves operating with  $(4N+2)\times(4N+2)$ matrices. The typical amount of computer memory limits the feasibly fast spectral simulation to $N\sim 10^2$. Remarkably, given reasonable values of the system parameters, the method we introduced in the previous section enables an analytical solution for the steady state and its spectral poles to be obtained for arbitrarily large values of $N$.  

The representation by Eq.~(\ref{0}) defines the decomposition $\rho^{(0,1)}\in\Lambda^{(0,1)}$ of the effective space into subspaces $\Lambda^{(0,1)}$ that satisfy the projection principle described in Appendix, A. Hence, the projections $\rho^{(0,1)}$ of the steady state are found independently from the equations 
\begin{equation}
\left[1-{\cal X}_0^2(\zeta)\right]\rho^{(0)}=\rho_{\rm th},\quad
\rho^{(1)}={\cal X}_0(\zeta)\rho^{(0)}.
\label{main}
\end{equation}
Using Eq.~(\ref{0}) and the physically reasonable condition $\Gamma\equiv\gamma_2+\Gamma_2+\Gamma_1/2\gg\gamma_1$, it can be shown that the first of Eqs.~(\ref{main}) is equivalent to the system
\begin{equation}
\begin{array}{c}
[\rho_0,I_+]=\eta\left(I_+\rho_z+\rho_zI_+-\rho_0I_+\right),\quad
{\rm Tr}\rho_0=1,\\[2mm]
2\gamma\rho_z=\left[{\cal L}(I_-)+{\cal L}(I_+)\right]\rho_z+I_-[I_+,\rho_0],\\[2mm]
{\displaystyle\gamma=\frac{\Gamma_1}{\gamma_1},\quad
\eta=\frac{\eta_0}{1+\zeta^2/\Gamma^2},\quad
\eta_0=\frac{4\Omega^2}{\gamma_1\Gamma}.}
\end{array}
\label{eq_rho0z}
\end{equation}
From the point of view of applications, the limit of large active/small passive longitudinal relaxation $\gamma\to\infty$ is particularly important. In this limit Eq.~(\ref{eq_rho0z}) enables the analytical solution valid for any $N$ (in the basis generated by the occupation numbers, denoting $\bar\eta=1+\eta$)
\begin{equation}
\rho_z=0,\quad
\rho_0=\frac{\eta\bar\eta^I}{\bar\eta^{N+1}-1}\sum_{n=-I}^I\bar\eta^{-n}\vert n\rangle\langle n\vert.
\label{exact0}
\end{equation}
The second of Eqs.~(\ref{main}) gives then 
\begin{equation}
\begin{array}{c}
{\displaystyle\rho_+=\frac{i\Omega}{\Gamma-i\zeta}\,\frac{\eta\bar\eta^I}{\bar\eta^{N+1}-1}
\sum_{n=-I}^I\sqrt{\lambda_n}\,\bar\eta^{-n}\vert n\rangle\langle n-1\vert,}\\[2mm]
\rho_-=\rho_+^\dagger,\quad
\lambda_n=(I-n+1)(I+n).
\end{array}
\label{exact1}
\end{equation}

The spectral character of the steady state is determined by the poles of the driven Green function that annihilate the denominators in Eqs.~(\ref{exact0}), (\ref{exact1})
$$
\zeta_r:\quad
\Gamma\pm i\zeta=0,\quad
\bar\eta^{N+1}-1=0\quad(\eta\not=0).
$$
This gives $N+1$ pairs of poles that are exactly calculated as ($m=1,\,\ldots,\,N$)
\begin{equation}
\zeta_0=\pm i\Gamma,\quad
\zeta_m=\pm i\Gamma\sqrt{1+\frac{\eta_0}{2}-i\frac{\eta_0}{2}\,{\rm cot}\frac{\pi m}{N+1}}.
\label{poles}
\end{equation}
The first pair $\zeta_0$ are the poles of the non-driven Green function. 

Eq.~(\ref{exact0}) can be used to estimate two important steady-state characteristics of the driven open quantum system, the polarization and self-correlation of the total $z$-component of the passive spin 
$$
\langle I_z\rangle={\rm Tr}\,\left(\rho_0I_z\right),\quad
\langle I^2_z\rangle={\rm Tr}\,\left(\rho_0I^2_z\right). 
$$
By virtue of Eq.~(\ref{exact0}), proceeding to continuous integral approximations, we obtain the expressions valid for arbitrarily large values of $N$
\begin{equation}
\begin{array}{c}
2\langle I_z\rangle/N=\lambda^{-1}-{\rm coth}\,\lambda,\quad
\lambda=I\ln\bar\eta,\\[2mm]
4\langle I_z\rangle/N^2=1+2\lambda^{-2}-2\lambda^{-1}{\rm coth}\,\lambda.
\end{array}
\label{II}
\end{equation}

\begin{figure}[t]
\begin{center}
\includegraphics[scale=0.37]{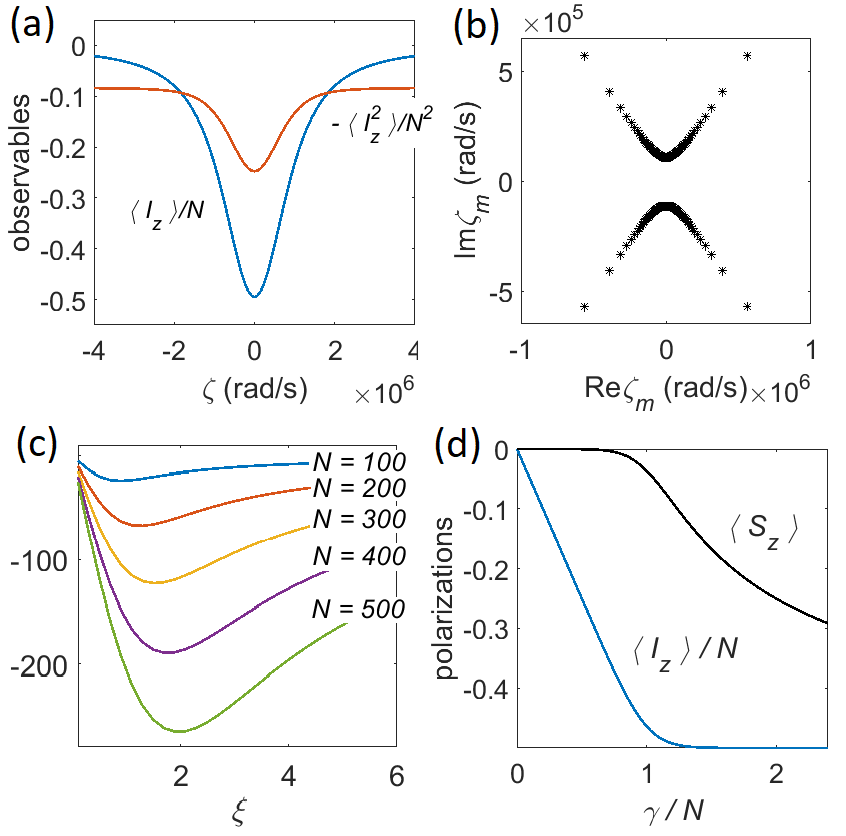}
\end{center}
\caption{
\label{1}
(a) Normalized polarization (blue) and self-correlation (red) of the total passive spin $z$-component as a function of the spectral parameter $\zeta$ for $N=10^3$ and $\eta_0=0.4$, $\Gamma=10^5\ {\rm rad/s}$.
(b) Poles of the ``absorption line'' on the complex plane for the same parameters as in panel (a).
(c) Total polarization $\xi\langle I_z\rangle$ of the passive ensemble at $\zeta=0$ as a function of the relative concentration $\xi$ of the active subsystems for different numbers $N$ for (all in rad/s) $\Gamma_2=10^6\xi^2$, $\Omega=10$, $\gamma_1=10^{-2}$, $\gamma_2=10^3$.   
(d) Polarization of the active subsystem (black) and normalized polarization of the passive ensemble (blue) in the limit $\eta_0\to\infty$ as a function of the relative relaxation parameter $\gamma$ for $\zeta=0$, $N=10^6$.
}
\end{figure} 

Numerical results by Eqs.~(\ref{II}) for a set of system parameters are plotted in FIG.~\ref{1}(a). It is evident that at $\zeta\sim 0$ the passive ensemble is almost fully polarized (population inverted) with $\langle I_z\rangle\sim-N/2$ that is accompanied with the creation of correlations between the passive subsystems. According to Eqs.~(\ref{poles}), the poles of the driven Green function are distributed on the complex plane symmetrically and densely around the origin $\zeta=0$, see FIG.~\ref{1}(b), leading to a single ``absorption line'' composed of $2N$ Lorentzian peaks at $\zeta={\rm Re}\,\zeta_m$ with widths $\vert{\rm Im}\,\zeta_m\vert$. It follows from this analysis that for large values of $\eta_0$ the spectral width grows linearly with $\Omega$, exactly as for a single passive subsystem $N=1$, while it grows nonlinearly as $\sim\sqrt{N}$ with the number of passive subsystems. Indeed, we have for large $N$: $\max\vert\zeta_m\vert=\vert\zeta_1\vert\sim\Gamma\sqrt{\eta_0(N+1)/2\pi}$.   

To illustrate the applicability of the method to optimization problems, consider now an ensemble of many active subsystems, each ``serving'' $N$ passive subsystems. Physically, the active transverse relaxation rate $\Gamma_2$ that influences the passive polarization is caused by active spin-spin interactions and grows quadratically with the spatial concentration $c$ of active subsystems. We can write $\Gamma_2=\Gamma_2^0\xi^2$ where $\xi=c^0/c$ is the dimensionless relative concentration with respect to some reference concentration $c^0$ and $\Gamma_2^0$ is the rate for $c=c^0$. For $\Gamma_2\gg\gamma_2+\Gamma_1/2$, the simulation by the first of Eqs.~(\ref{II}) implies that the total peak polarization of the passive ensemble $\xi\langle I_z\rangle$ at $\zeta=0$ has an active concentration optimum whose both location and peak value increase with $N$, FIG.~\ref{1}(c).  

Eqs.~(\ref{eq_rho0z}) enable us also to analyze the effect of the relative longitudinal relaxation described by the parameter $\gamma$. It follows from this equation that the active and passive polarizations are connected as $\langle S_z\rangle+\langle I_z\rangle/\gamma+1/2=0$.  Restricting to the limit $\eta_0\to\infty$, Eqs.~(\ref{eq_rho0z}) are resolved by a simple recurrency in the basis generated by the occupation numbers. As a result, for large $N$ the peak $\zeta=0$ dependence of the active and passive polarizations on the relative relaxation parameter $\gamma$ is well described as 
\begin{equation}
\begin{array}{c}
\langle S_z\rangle=0,\quad
\langle I_z\rangle=-\gamma/2,\quad
\gamma<N,\\[2mm]
\langle S_z\rangle=(\gamma/N-1)/2,\quad
\langle I_z\rangle=-N/2,\quad
\gamma>N.
\end{array}
\label{pt}
\end{equation}
For $N\gg 1$, the value $\gamma/N=1$ can be treated then as the critical value for the second order (continuous) phase transition between regimes dependent on and independent of the longitudinal relaxation: for $\gamma<N$ the active subsystem is fully saturated, for $\gamma>N$ the passive ensemble is fully polarized, FIG.~\ref{1}(d). This links our model to phase transitions predicted in the Dicke model \cite{lph-81,dc-78}.  

The details of the derivation of Eqs.~(\ref{eq_rho0z}), (\ref{exact0}), (\ref{exact1}), (\ref{poles}), (\ref{II}), (\ref{pt}) are given in Appendix, B. There we also discuss the situation where the passive subsystems are not necessarily identical giving links to the kinetic Monte Carlo algorithm \cite{bp-02,k-15,k-18}. 

{\bf\em Conclusion and acknowledgement.}
We have proposed a novel method of simulation of driven Markovian open quantum dynamics based on Green functions in spectral frequency domain. We demonstrated that the method is computationally highly efficient and opens up new ways in simulation, spectroscopy and optimization of many-body quantum dynamics in the realistically large-scale limit. This work was funded by the British Engineering and Physical Science Research Council (EPSRC) through grant EP/N03404X/1 to WK. 
\vskip 15mm

\section{Appendix}

{\bf\em A. ``Renormalization'' of perturbation series and links to time-domain Green functions and projection methods.}
Since the superoperator ${\cal X}_0(\zeta)$ satisfies its own characteristic equation, $\pi_0({\cal X}_0(\zeta))=0$, Eq.~(11) of the main text has the exact solution, polynomial in ${\cal X}_0(\zeta)$,
\begin{equation}
\rho=\bar\pi\left({\cal X}_0(\zeta)\right)\rho_{\rm th},\quad
\bar\pi(x)=\frac{\pi_0(1)-\pi_0(x)}{\pi_0(1)(1-x)}
\label{exact_rho}
\tag{A1}
\end{equation} 
where $\pi_0(x)$ is the characteristic polynomial of ${\cal X}_0(\zeta)$. Indeed, we have
$$
\begin{array}{c}
(1-{\cal X}_0)\rho=(1-{\cal X}_0)\bar\pi({\cal X}_0)\rho_{\rm th}=\\[2mm]
=\left[1-\pi_0({\cal X}_0)/\pi_0(1)\right]\rho_{\rm th}=\rho_{\rm th}.
\end{array}
$$
We have ${\cal X}_0={\cal G}_0{\cal P}$ where $\cal P$ is proportional to the commutation superoperator with the driving $P$. Hence, all operators commuting with $P$ belong to the zero subspace $V_0$ of the superoperator ${\cal X}_0$. It means that the latter is degenerate having the zero eigenvalue of multiplicity $m={\rm dim} V_0\ge N_0$ where $N_0$ is the dimension of the Hilbert space of the quantum problem. Indeed, all operators diagonal in the basis where $P$ is diagonal belong to $V_0$. The characteristic polynomial of ${\cal X}_0$ has then the form
$$
\pi_0(x)=x^m\pi'(x)
$$
where $\pi'(x)$ is a polynomial with nonzero roots. Eq.~(\ref{exact_rho}) implies then
$$
\bar\pi(x)=1+x+\ldots+x^{m-1}+x^m\bar\pi'(x)
$$ 
where $\bar\pi'(x)$ is a polynomial. It means that Eq.~(\ref{exact_rho}) describes a renormalization of the major coefficients of the generally divergent infinite perturbation series $1+x+x^2+\ldots$ in such way that the series is truncated to an always convergent polynomial expression.   

In the case where the superoperators ${\cal F}_0$, ${\cal H}_1$ commute $[{\cal F}_0,{\cal H}_1]=0$ the non-driven spectral Green function ${\cal G}_0(\zeta)$ can be written as a generalized Fourier transform
\begin{equation}
{\cal G}_0(\zeta)=-\int_{-\infty}^{+\infty}\bar{\cal G}_0(t)e^{-\zeta{\cal H}_1t}\,dt
\label{ft}
\tag{A2}
\end{equation} 
where $\bar{\cal G}_0(t)$ is the Green function of the inhomogeneous non-driven dynamical problem for $\zeta=0$
\begin{equation}
\dot\rho={\cal F}_0\rho+f.
\label{dyn}
\tag{A3}
\end{equation}
Indeed, for any bounded inhomogeneity $f$ the bounded solution to Eq.~(\ref{dyn}) is written as
$$
\begin{array}{c}
{\displaystyle\rho(t)=\int_{-\infty}^{+\infty}\bar{\cal G}_0(t-t')f(t')\,dt',}\\[4mm]
\bar{\cal G}_0(t)=e^{{\cal F}_0t},\quad
t\ge 0;\quad
\bar{\cal G}_0(t)=0,\quad
t<0.
\end{array}
$$
We have then
$$
\begin{array}{c}
{\displaystyle-\int_{-\infty}^{+\infty}\bar{\cal G}_0(t)e^{-\zeta{\cal H}_1t}\,dt=
-\int_0^{+\infty}e^{({\cal F}_0-\zeta{\cal H}_1)t}\,dt=}\\[4mm]
=\left({\cal F}_0-\zeta{\cal H}_1\right)^{-1}={\cal G}_0(\zeta).
\end{array}
$$
This implies that the magnitude $-{\cal G}_0(\zeta)\bar\rho(0)$ describes the generalized spectrum of the free thermal decay of the traceless part of an initial state $\rho(0)$. The superoperator $\bar G_0(t-t')$ of Eq.~(\ref{ft}) plays the role of the retarded Green function that describes the free irreversible decay of correlations between the initial state and the thermal equilibrium.

Suppose (that is typically the case) that the Liouville space $\Lambda=\Lambda^{(0)}+\Lambda^{(1)}$ is decomposed into two components that are invariant in the non-driven system and coupled by the driving, 
$$
\left({\cal F}_0-\zeta{\cal H}_1\right)\Lambda^{(0,1)}\subset\Lambda^{(0,1)},\quad
{\cal P}\Lambda^{(0,1)}\subset{\Lambda^{(1,0)}}
$$
with $\rho_{\rm th}\in\Lambda^{(0)}$. We obtain for the dynamics of the density operator projections $\rho^{(0,1)}\in\Lambda^{(0,1)}$
$$
\dot\rho^{(0)}={\cal A}^{(0)}\rho^{(0)}-{\cal P}\rho^{(1)},\quad
\dot\rho^{(1)}={\cal A}^{(1)}\rho^{(1)}-{\cal P}\rho^{(0)}
$$
where ${\cal A}^{(0,1)}$ are the restrictions of the non-driven superoperator to the subspaces $\Lambda^{(0,1)}$,
$$
{\cal A}^{(s)}=\left({\cal F}_0-\zeta{\cal H}_1\right)\big\vert_{\Lambda^{(s)}},\quad
s=0,\,1.
$$
If the non-driven dynamics in the subspace $\Lambda^{(1)}$ is much faster than its exchange with the subspace $\Lambda^{(0)}$, 
\begin{equation}
\vert{\rm eig}\,{\cal A}^{(1)}\vert\ll\Vert{\cal P}\Vert,
\label{ac}
\tag{A4}
\end{equation} 
then the subspace $\Lambda^{(1)}$ can be adiabatically eliminated. The dynamics in the subspace $\Lambda^{(1)}$ is well approximated by the quasi-equilibrium
$$
\rho^{(1)}=({\cal A}^{(1)})^{-1}{\cal P}\rho^{(0)}={\cal G}_0(\zeta){\cal P}\rho^{(0)}. 
$$
The dynamics of the projection to the subspace $\Lambda^{(0)}$ is well described then by the equation \cite{k-15,k-18}
\begin{equation}
\begin{array}{c}
\dot\rho^{(0)}=\left[{\cal A}^{(0)}-{\cal P}{\cal G}_0(\zeta){\cal P}\right]\rho^{(0)}=\\[2mm]
={\cal A}^{(0)}\left[1-{\cal X}^2_0(\zeta)\right]\rho^{(0)}.
\end{array}
\label{proj1}
\tag{A5}
\end{equation} 
The steady-state equation $\left(1-{\cal X}_0(\zeta)\right)\rho=\rho_{\rm th}$ implies that the steady-state projections $\rho^{(0,1)}$ satisfy the equations 
\begin{equation}
\left[1-{\cal X}^2_0(\zeta)\right]\rho^{(0)}=\rho_{\rm th},\quad
\rho^{(1)}={\cal X}_0(\zeta)\rho^{(0)}
\label{proj2}
\tag{A6}
\end{equation} 
regardless whether the adiabaticity condition (\ref{ac}) is fulfilled or not.

{\bf\em B. Mathematics of model example.}
The model Hamiltonian that we initially consider is built of one irradiated subsystem (called ``active'') described by the spin-1/2 angular momentum ${\bf S}$ and $N$ non-irradiated subsystems (that we call ``passive'') characterized by spin-1/2 angular momenta 
${\bf I}^{(k)}$, $1\le k\le N$. In the rotaing wave approximation, we have (in frequency units)
\begin{equation}
\begin{array}{c}
H=\omega_1S_x+\Delta S_z+\omega_II_z+H_{IS},\\[2mm]
{\displaystyle H_{IS}=\frac{1}{2}\sum_{k=1}^N\left[A_kI^{(k)}_++A^*_kI^{(k)}_-\right]S_z.}
\end{array}
\label{Hm}
\tag{B1}
\end{equation}
It is assumed that the active and passive level separation frequencies satisfy the condition $\omega_S\gg\omega_I$ and the effective irradiation acts along the $x$-axis orthogonal to the quantization $z$-axis and has the strength $\omega_1$ and frequency $\omega_0\gg\omega_0$. Then $\Delta=\omega_S-\omega_0$ characterizes the offset of the irradiation frequency from the level separation frequency of the active subsystem, while the level separation frequency $\omega_I$ of the passive subsystems remains unchanged. The term $H_{IS}$ describes the interactions of the passive subsystems with the active subsystem that take into account single-quantum passive spin coherences $I^{(k)}_\pm=I^{(k)}_x+iI^{(k)}_y$ and the active-passive interaction strengths $A_k$. This term is the only coherent term of the dipole-dipole interactions that commutes with $S_z$ and is preserved in the rotating wave approximation. By a suitable rotation of the transverse spin components $I^{(k)}_\pm\to I^{(k)}_\pm e^{\pm i\phi_k}$ we can always achieve that the interaction strengths $A_k$ coincide with their absolute values. We can write then
\begin{equation}
\begin{array}{c}
2H_{IS}=A(V_++V_-)S_z,\quad
A=\sum_k\vert A_k\vert/N,\\[2mm]
V_\pm=\sum_ka_kI^{(k)}_\pm,\quad
a_k=\vert A_k\vert/A
\end{array}
\label{V}
\tag{B2}
\end{equation}
where $A$ is the average absolute value of the active-passive interaction strengths and the factors $a_k\ge 0$ characterize the contributions of the passive subsystems to the active-passive interactions. 

To describe the dissipation, we assume that the spin interaction strengths $\vert A_k\vert\ll\omega_{I,S}$ are much smaller than the level separation frequencies. Then the thermal equilibrium is well described by the Boltzmann distribution of the energies along the quantization axis
$$
\begin{array}{c}
\rho_{\rm th}=Z^{-1}\exp\left[-\beta(\omega_SS_z+\omega_II_z)\right]=\\[2mm]
{\displaystyle=\left(\frac{1}{2}-p_SS_z\right)\prod_{k=1}^N\left(\frac{1}{2}-p_II^{(k)}_z\right),}\\[4mm]
{\displaystyle p_S=\tanh\,\frac{\beta\omega_S}{2},\quad
p_I=\tanh\,\frac{\beta\omega_I}{2},\quad
\beta=\hbar/kT.}
\end{array}
$$ 
In the case where the active subsystem is ``cold'' and the passive subsystems are ``hot'' with respect to the thermal energy,
$\hbar\omega_I\ll kT\ll\hbar\omega_S$, the thermal equilibrium is approximated as $\rho_{\rm th}=2^{-N}(1/2-S_z)$ where the active subsystem is in the ground state $p_S\sim 1$ while the passive subsystems have all equaly populated levels $p_I\sim 0$. The typical Lindblad dissipator preserves the thermal equilibrium and has the form
\begin{equation}
\begin{array}{c}
{\cal D}={\cal D}_S+{\cal D}_I,\quad
{\cal D}_S=\Gamma_1{\cal L}(S_-)+2\Gamma_2{\cal L}(S_z),\\[2mm]
{\displaystyle{\cal D}_I=\frac{\gamma_1}{2}\left[{\cal L}(V_+)+{\cal L}(V_-)\right]+2\gamma_2{\cal L}(I_z).}
\end{array}
\label{Dm}
\tag{B3}
\end{equation}       
Here $\Gamma_1,\,\Gamma_2,\,\gamma_1,\,\gamma_2>0$ are the effective active and passive longitudinal and transverse relaxation rates, $V_\pm$ are dimensionless jump operators given by Eq.~(\ref{V}). We also take into account the passive transverse relaxation in the simplest collective average form. 

Master equations with the Hamiltonian and dissipative parts in the form of Eqs.~(\ref{Hm}), (\ref{Dm}) are met, for example, in quantum optics where they describe optically irradiated unlike 2-level atomic systems \cite{s-98,a-74}. Eqs.~(\ref{Hm}), (\ref{Dm}) are typical also for dynamic nuclear polarization where they describe microwave irradiated electron-nuclear paramagnetic systems \cite{a-61,wb-07}. It is important for our study that in both cases the passive longitudinal relaxation is relativey slow, so that the following condition is well satisfied
\begin{equation}
\gamma_1\ll\Gamma=\gamma_2+\frac{\Gamma_1}{2}+\Gamma_2.
\label{gam1}
\tag{B4}
\end{equation} 
In optics this is because $\omega_I\ll\omega_S$ and so $\gamma_1\sim(\omega_I/\omega_S)^3\Gamma_1$ as follows from the spontaneous emission theory \cite{s-98,a-74}. In dynamic nuclear polarization, Eq.~(\ref{gam1}) is satisfied in the high-field 
low-temperature limit where $\gamma_1\sim (A/\omega_I)^2(1-p_S^2)\Gamma_1$ in accordance with the theory of nuclear relaxation by paramagnetic impurities \cite{b-49,ag-78}. Hence, condition (\ref{gam1}) holds independently of the transverse relaxation rates $\gamma_2,\,\Gamma_2$.      

Our next step is to consider the ``solid effect'' resonance where the active frequency offset is comparable to the passive frequency, $\Delta\sim\omega_I$. In this case, the active spin flips/flops are ``synchronized'' with the passive spin flops/flips. Using the adiabatic elimination method \cite{k-15,kSE-12}, the Hamiltonian (\ref{V}) is transformed to a 2-spin flip-flop Hamiltonian
\begin{equation}
H_{\rm eff}=\zeta S_z+\Omega(V_+S_-+V_-S_+),\quad
\Omega=\frac{\omega_1A}{4\omega_I}.
\label{Heff}
\tag{B5}
\end{equation}     
Here $\zeta$ is the resonance offset $\zeta=\Delta-\omega_I$. The dissipator (\ref{Dm}) remains unchanged. Similarly, the case $\Delta\sim-\omega_I$ leads to an effective 2-spin flip-flip Hamiltonian \cite{k-15,kSE-12}.  

The master equation with the Hamiltonian and dissipative parts defined by Eqs.~(\ref{Heff}), (\ref{Dm}) preserves the subspace $\Lambda_0$ of zero-quantum coherences, $[I_z+S_z,\Lambda_0]=0$. Since $\rho_{\rm th}\in\Lambda_0$, the driven dynamics and the steady state are closed in $\Lambda_0$. The ${\dim}\,\Lambda_0=[2(N+1)]!/[(N+1)!]^2$ exponentially grows with $N$. The typical volume of computer memory limits the feasibly fast spectral simulation within the full master equation to $N<10$ that is far from a physically realistic assumption. Remarkably, the method we introduced in the main text enables to extend the feasible number of the passive subsystems to $N\sim 10^3$. The ``mean-field'' approximation with $a_k=1$ enables an analytical solution for the steady state and its spectral poles to be obtained for arbitrarily large values of $N$.  

In the notations of the main text,
\begin{equation}
\begin{array}{c}
H_0=0,\quad
H_1=S_z,\\[2mm]
P=\Omega(P_++P_-),\quad
P_\pm=V_\mp S_\pm.
\end{array}
\label{P}
\tag{B6}
\end{equation}
The density operator admits the decomposition
\begin{equation}
\begin{array}{c}
\rho=\rho^{(0)}+\rho^{(1)},\quad
\rho^{(1)}=\rho_-S_++\rho_+S_-,\\[2mm]
\rho^{(0)}=\rho_0\left(1/2-S_z\right)+2\rho_zS_z
\end{array}
\label{0}
\tag{B7}
\end{equation} 
with $\rho_{0,z,\pm}$ containing only the passive spin components. We have $\rho^{(0,1)}\in\Lambda^{(0,1)}$, 
$\rho_{\rm th}\in\Lambda^{(0)}$ where the subspaces $\Lambda^{(0,1)}$ built of zero-quantum and single-quantum coherences of the active subsystem satisfy the conditions of the projection method described in the previous section. Eqs.~(\ref{proj1}), (\ref{proj2}) imply that the projections $\rho^{(0,1)}$ of the steady state are found independently from the equations 
\begin{equation}
\left[1-{\cal X}_0^2(\zeta)\right]\rho^{(0)}=\rho_{\rm th},\quad
\rho^{(1)}={\cal X}_0(\zeta)\rho^{(0)}.
\label{main}
\tag{B8}
\end{equation}
Applying to both sides of the first of Eqs.~(\ref{main}) the superoperator ${\cal F}_0-\zeta{\cal H}_1$ and using Eq.~(10) of the main text, we see that the first of Eqs.~(\ref{main}) is equivalent to the equation
\begin{equation}
\left[{\cal F}_0-\zeta{\cal H}_1-{\cal P}{\cal G}_0(\zeta){\cal P}\right]\rho^{(0)}=0,\quad
{\rm Tr}\rho^{(0)}=1.
\label{e1}
\tag{B9}
\end{equation}

For any operator $\rho^{(0)}\in\Lambda^{(0)}$, we obtain
$$
\begin{array}{c}
\left[{\cal F}_0-\zeta{\cal H}_1\right]\rho^{(0)}={\cal D}^{(1)}\rho^{(0)}\in\Lambda^{(0)},\\[2mm]
[P_\pm,\rho^{(0)}]=\rho_\mp S_\pm\in\Lambda^{(1)}
\end{array}
$$
with $\rho_\pm$ containing only passive spin components and ${\cal D}^{(1)}$ denoting the longitudinal part of the dissipator in Eq.~(\ref{Dm}). For any operators $\rho_\pm$ we have
$$
\left[{\cal F}_0-\zeta{\cal H}_1\right](\rho_\mp S_\pm)=\left[{\cal D}^{(1)}_I\mp i\zeta-\Gamma\right]\rho_\mp S_\pm\in\Lambda^{(1)}
$$
where ${\cal D}^{(1)}_I$ is the longitudinal part of the passive dissipator in Eq.~(\ref{Dm}) and $\Gamma$ is defined in Eq.~(\ref{gam1}). The latter implies that the passive longitudinal relaxation makes a negligible contribution to the dynamics in the subspace $\Lambda^{(1)}$. As a result, by virtue of Eq.~(10) of the main text that defines the non-driven Green function,
$$
\begin{array}{c}
{\cal G}_0(\zeta)(\rho_\mp S_\pm)=\left[{\cal F}_0-\zeta{\cal H}_1\right]^{-1}(\rho_\mp S_\pm)=\\[2mm]
=-\left(\Gamma\pm i\zeta\right)^{-1}\rho_\mp S_\pm.
\end{array}
$$  
Since $[P_\pm,\rho_\mp S_\pm]=0$, we obtain
$$
\begin{array}{c}
{\cal P}{\cal G}_0(\zeta){\cal P}\rho^{(0)}=\\[2mm]
{\displaystyle=\Omega^2\left(\frac{[P_-,[P_+,\rho^{(0)}]]}{\Gamma+i\zeta}+
\frac{[P_+,[P_-,\rho^{(0)}]]}{\Gamma-i\zeta}\right)=}\\[4mm]
{\displaystyle=-\frac{\Omega^2}{\Gamma^2+\zeta^2}\left(2\Gamma{\cal L}(P_++P_-)\rho^{(0)}-
i\zeta[P_0,\rho^{(0)}]\right),}\\[4mm]
P_0=[P_+,P_-].
\end{array}
$$
Eq.~(\ref{e1}) can be rewritten then as
\begin{equation}
\begin{array}{c}
-i[H_0,\rho^{(0)}]+{\cal D}_0\rho^{(0)}=0,\quad
{\rm Tr}\rho^{(0)}=1,\\[2mm]
{\displaystyle H_0=\frac{f\omega P_0}{1+f^2},\quad
{\cal D}_0={\cal D}^{(1)}+\frac{2\omega{\cal L}(P_++P_-)}{1+f^2},}\\[3mm]
\omega=\Omega^2/\Gamma,\quad
f=\zeta/\Gamma
\end{array}
\label{e2}
\tag{B10}
\end{equation}
It is seen that the right-hand sides of Eqs.~(\ref{e2}) are fully determined by the dimensionless magnitudes $a_k$ that participate in the expressions for $V_\pm$, $P_\pm$ in Eqs.~(\ref{V}), (\ref{P}) and four physical parameters: the dimensionless spectral parameter $f$, the effective irradiation strength $\omega$ and the longitudinal relaxation rates of the passive and active subsystems $\gamma_1$, $\Gamma_1$. Note also that ${\cal L}(P_++P_-)={\cal L}(P_+)+{\cal L}(P_-)$.

It follows from Eq.~(\ref{e2}) that under the condition $\Gamma_1\gg\gamma_1$, independently on values taken by $f,\,\omega$, the steady state is well approximated by an operator that annihilates the active longitudinal dissipation. Indeed, for values of the magnitudes $\omega/(1+f^2)$, $f\omega/(1+f^2)$ much smaller than $\Gamma_1$, the active longitudinal dissipation dominates in Eq.~(\ref{e2}) and so the solution should annihilate it. For values of  $\omega/(1+f^2)$, $f\omega/(1+f^2)$ much larger than $\Gamma_1$, the relaxation processes are much slower than the active-passive exchange. In the first approximation the solution should commute with the operators $P_\pm$, i.e., should be a function of the total spin $z$-component, $\rho^{(0)}=F(I_z+S_z)=F_1(I_z)+F_2(I_z)S_z$. The second major term of the asymptotics is given by the active longitudinal relaxation, so the solution should again annihilate it. It is sufficient then to consider the projection of Eq.~(\ref{e2}) to the subspace of operators annihilating the active longitudinal relaxation, $\rho^{(0)}=\rho_0(1/2-S_z)$ where $\rho_0$ depends only on passive spin components and satisfies the equation      
\begin{equation}
\begin{array}{c}
{\displaystyle\frac{\gamma_1}{2}{\cal L}(V_+)\rho_0+\left[\frac{\gamma_1}{2}+\chi(f)\right]{\cal L}(V_-)\rho_0-}\\[4mm]
{\displaystyle-if\chi(f)[V_-V_+,\rho_0]=0,\quad
{\rm Tr}\rho_0=1.}
\end{array}
\label{epr}
\tag{B11}
\end{equation}   

Eq.~(\ref{epr}) shows that in the limit $\zeta\to\infty$ terms with $\chi(f)$ vanish, so the thermal equilibrium is preserved, 
$\rho_0\sim 2^{-N}$ and no polarization of the passive subsystems is created. The asymptotics $\zeta=0$, $\omega\gg\gamma_1$ leads to the equation ${\cal L}(V_-)\rho_0=0$, ${\rm Tr}\rho_0=1$ that has the solution 
$\displaystyle\rho_0=\prod_k\left(\frac{1}{2}-I^{(k)}_z\right)$ that corresponds to the fully polarized (population inverted) state of the passive ensemble,
\begin{equation}
\langle I_z\rangle={\rm Tr}\left(\rho_0\sum_kI^{(k)}_z\right)=-N/2.
\label{mp}
\tag{B12}
\end{equation}
The intermediate values of $\zeta$ generate a symmetric ``absorption line'' $\langle I_z\rangle(\zeta)$ that is zero at $\zeta\to\infty$ and has a peak of the maximal polarization (\ref{mp}) at $\zeta=0$. 

The shape and the width of the absorption line can be estimated in the ``mean-field'' approximation obtained by setting the magnitudes $a_k$ in Eq.~(\ref{V}) to be all equal $a_k=1$. In Eq.~(\ref{epr}) we obtain then $V_\pm=I_\pm$, the passive subsystems become identical and the dynamics is fully defined by the components of the total passive spin. This case is simplified by representation of the passive ensemble by a single angular momentum $\bf I$ with the spin quantum number $I=N/2$ similar to the Dicke model \cite{d-54,bzpp-69,g-11,lph-81,dc-78}. The corresponding occupation numbers are defined as $n=(n_+-n_-)/2=-I,\,-I+1,\,\ldots,\,I$ where $n_\pm$ are the numbers of the passive subsystems in the excited/ground state. In this ansatz \cite{r-95}
\begin{equation}
\begin{array}{c}
{\displaystyle I_z=\sum_{n=-I}^In\vert n\rangle\langle n\vert,\
I_+=\sum_{n=-I}^I\sqrt{\lambda_n}\,\vert n\rangle\langle n-1\vert,}\\[4mm]
I_-=I^\dagger_+,\quad
\lambda_n=(I-n+1)(I+n).
\end{array}
\label{I}
\tag{B13}
\end{equation}  
The operators $V_-V_+=I_-I_+$ and $\rho_0$ are diagonal in the basis generated by the occupation numbers, so the contribution of the Hamiltonian part of Eq.~(\ref{epr}) is zero. Denoting $\rho_0=\sum_nu_n\vert n\rangle\langle n\vert$, we obtain by virtue of Eq.~(\ref{I})
$$
\begin{array}{c}
{\cal L}(I_+)\rho_0=\sum_n\lambda_{n+1}u_n\left(\vert n+1\rangle\langle n+1\vert-\vert n\rangle\langle n\vert\right),\\[2mm]
{\cal L}(I_-)\rho_0=\sum_n\lambda_nu_n\left(\vert n-1\rangle\langle n-1\vert-\vert n\rangle\langle n\vert\right)
\end{array}
$$
that leads to the exact analytical solution to Eq.~(\ref{epr}) that is valid for any $N$ and given by Eq.~(19) of the main text. The second of Eqs.~(\ref{main}) leads then to Eq.~(20) of the main text.

The poles of the driven Green function annihilate the denominators in Eqs.~(19), (20)
$$
\zeta_r:\quad
\Gamma\pm i\zeta=0,\quad
1-\bar\eta^{N+1}=0\quad(\eta\not=0).
$$
We have
$$
\bar\eta=\exp\left(\frac{2im\pi}{N+1}\right),\quad
m=1,\,2,\,\ldots,\,N,
$$ 
which after simple algebra gives Eqs.~(21) of the main text. 

Eqs.~(22) of the main text describing the polarization and self-correlation of the total $z$-component of the passive spin are obtain by proceeding from the discrete set of the occupation numbers to the continuous interval 
$x=n/I\in[-1,1]$ and replacing discrete summations over $n$ by integrals with respect to $x$ using the smallness of the discrete step $1/I$ in the interval $[-1,1]$ \cite{bzpp-69}. For example,     
$$
\begin{array}{c}
{\displaystyle\langle I_z\rangle={\rm Tr}\,\left(\rho_0I_z\right)=c\sum_{n=-I}^In\bar\eta^{-n}=
cI^2\int_{-1}^1xe^{-\lambda x}\,dx=}\\[4mm]
{\displaystyle=\frac{2I^2}{\lambda^2}\left(\sinh\lambda-\lambda\cosh\lambda\right),\quad
\lambda=I\ln\bar\eta,}\\[4mm]
{\displaystyle c=\left(\sum_{n=-1}^I\bar\eta^{-n}\right)^{-1}=\left(I\int_{-1}^1e^{-\lambda x}\,dx\right)^{-1}=\frac{2I}{\lambda}\sinh\lambda}
\end{array}
$$
leading to $\langle I_z\rangle=I(\lambda^{-1}-\coth\lambda)$. Similarly the second moment $\langle I_z^2\rangle$ is calculated. 

To obtain Eqs.~(23) of the main text, we first rewrite the firsts of Eqs.~(\ref{main}) in terms of the decomposition of Eq.~(\ref{0}) applying the ``mean-field'' approximation $V_\pm=I_\pm$. After simple algebra this gives Eqs.~(18) of the main text. Denoting
$$
\rho_0=\sum_{n=-I}^Iu_n\vert n\rangle\langle n\vert,\quad
\rho_z=\sum_{n=-I}^Iv_n\vert n\rangle\langle n\vert
$$   
and using Eqs.~(\ref{I}), we come to the decoupled recurrency
\begin{equation}
\begin{array}{c}
{\displaystyle\left(\frac{2\gamma+\lambda_n}{\lambda_{n+1}}+1\right)v_n=
\left(\frac{\lambda_n}{\lambda_{n+1}}+1\right)v_{n-1},}\\[3mm]
n=-I+1,\,\ldots,\,I-1,\\[2mm]
{\displaystyle\left(\frac{2\gamma}{N}+1\right)v_I=v_{I-1},}\\[3mm]
u_{n+1}=v_{n+1}+v_n,\quad
n=-I,\,\ldots,\,I-1,\\[2mm]
u_{-I}+u_{-I+1}+\ldots+u_I=1.
\end{array}
\label{uv}
\tag{B14}
\end{equation}
valid in the limit $\eta\to\infty$. Solving Eqs.~(\ref{uv}) numerically for large $N$, we obtain FIG.~1(d) and Eqs.~(23) of the main text.  

Note finally that Eqs.~(\ref{e2}), (\ref{epr}) are both of the Lindblad form. They can be treated by unravelling in Hilbert space using the kinetic Monte Carlo method \cite{bp-02}. Here only four jump operators $V_\pm$, $P_\pm$ are involved in the computation scheme. In the case where the transverse relaxation of the passive ensemble is dominated by the individual dephasing mechanism 
$2\gamma_2\sum_k{\cal L}(I^{(k)}_z)$ with the strong rate $\gamma_2\gg\gamma_1$, the dynamics remains closed in the subspace spanned by $z$-components $I^{(k)}_z$ of the passive spins. In this subspace the collective Lindbald terms ${\cal L}(V_\pm)$, ${\cal L}(P_\pm)$ are split into sums of individual terms $\sum_k{\cal L}(I^{(k)}_\pm)$, $\sum_k{\cal L}(I^{(k)}_\mp S_\pm)$. In the latter case, the kinetic Monte Carlo scheme is reduced to sign permutations in a subsequence of $N+1$ symbols. This extends the feasible number of passive subsystems from $N\sim 10$ to $N\sim 10^3$, see Refs.~\cite{k-15,k-18} for details.

\end{document}